\documentclass[9pt]{article}
\usepackage{spconf,amsmath,epsfig}

\usepackage{amsmath}
\usepackage{amsfonts}
\usepackage{latexsym}
\usepackage{graphicx}
\usepackage{color}

\usepackage{amssymb,mathrsfs}
\usepackage{float}
\usepackage{subfigure}
\usepackage{slashbox}

\title{A SPECTRAL ANALYSIS APPROACH FOR GAUSSIAN MIXTURE ESTIMATION}

\name{Nicolas Paul, Michel Terre, Luc Fety} 
\address{Electronique et Communications, CNAM, 292 rue Saint-Martin, 75141 Paris Cedex 3, France \\ 
Telephone: 33 1 40 27 25 67, Fax: 33 1 40 27 24 81, Email: nicolas.paul@cnam.fr}

\begin{document}
%
\maketitle
%
\begin{abstract}
This paper deals with the estimation of one-dimensional Gaussian mixture. Given a set of observations of a $K$-component Gaussian mixture, we focus on the estimation of the component expectations. The number of components is supposed to be known. Our method is based on a spectral analysis of the estimated first characteristic function. We construct a Toeplitz matrix $\mathbf{R}_M$ with $2M-1$ estimated samples of the first characteristic function and show that the mixture component expectations can be derived from the eigenvector decomposition of $\mathbf{R}_M$. Simulations illustrate the performance of our algorithm on several configurations of a six-component Gaussian mixture. In the investigated scenarios the proposed method outperforms the Expectation-Maximization algorithm. 
\end{abstract}


\section{introduction}

In this paper we deal with Gaussian mixture estimation. Given a set of one-dimensional observations originating from $K$ possible Gaussian components, we focus on the estimation of the component expectations. The number of components is supposed to be known, and the component expectations are supposed to be all different.

One method consists in estimating a sampling of the observations probability density function (pdf), a mixture of $K$ pdf, by associating a kernel to each observation and adding the contribution of all the kernels \cite{Parzen}. A search of the pdf modes then leads to the component expectations. The drawback of such a method is that it requires the selection of extra-parameters (kernel design, sampling intervals). Furthermore, the final mode search algorithm might fail because of spurious local maxima in the estimated pdf.

An alternative method consists in using the Expectation-Maximization (EM) algorithm \cite{Dempster}. Each EM iteration consists of two steps. The Expectation step estimates the probability for each observation to come from each mixture component. Then, during the Maximization step, these estimated probabilities are used to update the estimation of the mixture parameters. This procedure converges to a local maximum of the likelihood. The main drawback of the EM algorithm is the potential convergence to some local non-global maxima of the likelihood. Some solutions consist, for instance, in using smart initializations or stochastic optimization \cite{McLachlan}.

In this contribution we propose a new approach based on a spectral analysis of the first characteristic function (CF). We define a Toeplitz matrix $\mathbf{R}_M$ with $2M-1$ estimated samples of the CF and show that the mixture component expectations can be estimated from an eigenvector decomposition of $\mathbf{R}_M$. The proposed method is strongly inspired from the \textit{mu}ltiple \textit{si}gnal \textit{c}lassification algorithm MUSIC \cite{Schmidt} which aims at estimating the frequencies in a sum of sinusoids. The paper is organized as follow: In section 2 the observation model is presented and an analytical expression of the CF of a Gaussian mixture is given. In section 3 the matrix $\mathbf{R}_M$ is defined and some properties of $\mathbf{R}_M$ are described. Section 4 then presents the complete estimation algorithm. Section 5 illustrates the estimation performances on a six-component Gaussian mixture with different configurations. Conclusions are finally given in section 6, as well as perspectives for using the proposed method to estimate the number of components in a mixture.

\section{Gaussian mixture}

\subsection{Probability density function (pdf)}
\label{secPDF}

Let $\{ p_k \}_{k\in\{1 \cdots K\}}$ be a set of $K$ positive mixing weights that sum up to one. The multimodal pdf of the random observable variable $Z$ is a finite mixture given by: 
\begin{equation}
f_{Z}(z)\stackrel{\Delta}{=}\sum_{k=1}^{K}{ p_k g(z,a_k,\sigma_k) },
\label{mixture}
\end{equation}
\noindent where $g(z,a_k,\sigma_k)$ is the Gaussian pdf given by: $g(z,a_k,\sigma_k)=\frac{1}{\sqrt{2\pi}\sigma_k}\text{exp}\left(-\frac{(z-a_k)^2}{2\sigma_k^2} \right)$ and $a_k$ and $\sigma_k$ are respectively the expectation and the standard deviation of component $k$. Given a set of $N$ observed realizations $\{z_n\}_{n\in\{1,\cdots,N\}}$ of $Z$ we focus on the estimation of the $K$ component expectations $\{a_k\}_{k\in\{1,\cdots,K\}}$. Our proposal is mainly based on the estimated first characteristic function (CF) of the mixture.

\subsection{First characteristic function (CF)}
\label{secCF}

In general, the CF of a random variable $X$ is defined by:
\begin{equation}
\phi_X(t)\stackrel{\Delta}{=}E_X\{e^{itX}\}, \qquad t\in\mathbb{R},
\end{equation}
where $E_X\{\cdot\}$ is the mathematical expectation with respect to the pdf of $X$. For instance, the CF of a Gaussian random variable $X$ with pdf $g(x,a,\sigma)$ is given by \cite{Picinbono}:
\begin{eqnarray}
\phi_X(t)
&=&\int_{x=-\infty}^{\infty}{e^{itx}g(x,a,\sigma)\text{d}x} \\
&=&e^{-\frac{\sigma^2t^2}{2}}e^{iat}. 
\end{eqnarray}
Consequently the CF of the random variable $Z$ with the pdf described in \eqref{mixture} is given by:
\begin{eqnarray}
\phi_Z(t)
&=&\sum_{k=1}^{K}p_k\int_{z=-\infty}^{\infty}{e^{itz}g(z,a_k,\sigma_k)\text{d}z} \\
&=&\sum_{k=1}^{K}{p_k e^{\frac{-\sigma_k^2t^2}{2}} e^{ita_k}}. 
\label{mixtureContinuousCF} 
\end{eqnarray}
Now let $\phi_m$ be the sampled version of $\phi_Z(t)$ with a sampling period $T_e$. According to \eqref{mixtureContinuousCF}, we have:
\begin{eqnarray}
\phi_m
&\stackrel{\Delta}{=}&\phi_Z(mT_e), \qquad m\in \mathbb{Z} \label{CFsampledDef} \\
&=&\sum_{k=1}^{K}{p_k\alpha_{k,m} w_k^m}, 
\label{CFsampled}
\end{eqnarray}
where $w_k$ and $\alpha_{k,m}$ are defined by:
\begin{eqnarray}
w_k&\stackrel{\Delta}{=}&e^{ia_kT_e} \label{defAk} \\
\alpha_{k,m}&\stackrel{\Delta}{=}&e^{-\frac{\sigma_k^2(mT_e)^2}{2}}.
\label{defAlpha}
\end{eqnarray}
In practical situation, $\phi_m$ can be estimated from a set of $N$ observations $\{z_n\}_{n=1,\cdots,N}$ using:
\begin{equation}
\hat{\phi}_m=\frac{1}{N}\sum_{n=1}^{N}e^{iz_n(mTe)}
\label{estimationCF}
\end{equation}
In section \ref{DefPropRM} we will show how the $\{w_k\}_{k=1,\cdots,K}$ defined in \eqref{defAk} can be estimated from the sampled CF. Once the $\{w_k\}_{k=1,\cdots,K}$ are estimated, the $\{a_k\}_{k=1,\cdots,K}$ can be obtained without ambiguity if the sampling period $T_e$ is less than $\frac{2\pi}{\text{max}\{z_n\}-\text{min}\{z_n\}}$: If we for instance choose:
\begin{equation}
T_e=\frac{2\pi}{2(\text{max}\{z_n\}-\text{min}\{z_n\})},
\label{Te}
\end{equation}
then $\frac{2\pi}{T_e}=2(\text{max}\{z_n\}-\text{min}\{z_n\})$. Since $\text{min}\{z_n\}\leq a_k\leq\ \text{max}\{z_n\}$
there is exactly one integer $l_{w_k}$ such as: 
\begin{equation}
\frac{\text{angle}(w_k)}{T_e}+l_{w_k}\frac{2\pi}{T_e} \in [\text{min}\{z_n\} \quad \text{max}\{z_n\}],
\end{equation}
and we have:
\begin{equation}
a_k=\frac{\text{angle}(w_k)}{T_e}+ l_{w_k}\frac{2\pi}{T_e}, \qquad k=1,\cdots,K.
\label{ak_from_wk}
\end{equation}
Furthermore, if $T_e$ verifies \eqref{Te}, and since the $\{a_k\}_{k=1,\cdots,K}$ are supposed to be all different, then the $\{w_k\}_{k=1,\cdots,K}$ are also all different. This will be used in section \ref{DefPropRM}.

\section{Definition and properties of $\mathbf{R}_M$}
\label{DefPropRM}
Let $\mathbf{R}_M=(r_{jl})_{j,l=1,\cdots,M} \in \mathbb{C}^{M\times M}$ be the Toeplitz matrix with the following elements:
\begin{equation}
r_{jl}=\phi_{l-j}, \qquad j,l=1,\cdots,M
\label{defRM}
\end{equation}
\noindent where $\phi_m$ has been defined in \eqref{CFsampledDef}. Note that $\phi_{-m}=\phi_{m}^*$ so $\mathbf{R}_M$ is a Hermitian matrix and one only has to compute $M$ samples of the CF to build $\mathbf{R}_M$. Including \eqref{CFsampled} into \eqref{defRM}:
\begin{eqnarray}
r_{jl}
&=&\sum_{k=1}^{K}{p_k\alpha_{k,l-j} w_k^{l-j}} \\
&=&\sum_{k=1}^{K}{p_k w_k^{l-j}} + \sum_{k=1}^{K}{p_k(\alpha_{k,l-j}-1)w_k^{l-j}} \\
&=&\sum_{k=1}^{K}{{w_k^*}^{j-1}p_kw_k^{l-1}} + \sum_{k=1}^{K}{p_k(\alpha_{k,l-j}-1)w_k^{l-j}}, \notag \\
\label{rjlForme}
\end{eqnarray}
where in \eqref{rjlForme} we used that $w_k^{l-j}={w_k^*}^{j}w_k^{l}={w_k^*}^{j-1}w_k^{l-1}$ since ${w_k^*}^{-1}w_k^{-1}=1$. A consequence of \eqref{rjlForme} is that $\mathbf{R}_M$ can be expressed as the sum of a "signal" matrix $\mathbf{S}_M$ and a "perturbation" matrix $\mathbf{P}_M$:
\begin{equation}
\mathbf{R}_M=\mathbf{S}_M+\mathbf{P}_M,
\end{equation}
where the "signal" matrix $\mathbf{S}_M$ is given by:
\begin{eqnarray}
\mathbf{S}_M&\stackrel{\Delta}{=}&(\mathbf{w}_1,\cdots,\mathbf{w}_K)\mathbf{D} 
\begin{pmatrix}
\mathbf{w}_1^H \\
\vdots \\
\mathbf{w}_K^H 
\end{pmatrix} \in \mathbb{C}^{M\times M} \label{defSM} \\
\mathbf{w}_k&\stackrel{\Delta}{=}&(1,w_k^1,\cdots,w_k^{M-1})^H \in \mathbb{C}^M \label{def_wk} \\
\mathbf{D}&\stackrel{\Delta}{=}&\text{diag}(p_1,\cdots,p_K) \in \mathbb{R}^{K\times K} 
\end{eqnarray}
and the "perturbation" matrix $\mathbf{P}_M$ is given by: 
\begin{equation}
\mathbf{P}_M\stackrel{\Delta}{=}\sum_{k=1}^{K}{p_k\left((\alpha_{k,l-j}-1)w_k^{l-j}\right)_{l,j=1,\cdots,M}} \in \mathbb{C}^{M\times M}. 
\label{defPM}
\end{equation}
The "signal" matrix $\mathbf{S}_M$ is a well-known matrix in the spectral analysis community. It is the auto-correlation matrix of a received sum of $K$ sinusoids with angular frequencies $a_k$ and power $p_k$. High resolution algorithm such as MUSIC \cite{Schmidt} estimate $\mathbf{S}_M$ from some (potentially corrupted) signal samples then estimate the sinusoid frequencies from its eigenvector decomposition. Indeed, since the $w_k$ are all different (section \ref{secCF}), one can show that the rank of $\mathbf{S}_M$ is equal to $K$ and that the signal vectors $\mathbf{w}_k$ defined in \eqref{def_wk} are orthogonal to any vector of the kernel of $\mathbf{S}_M$ \cite{Haykin}. Consequently, if $\mathbf{V}\stackrel{\Delta}{=}(\mathbf{v}_{K+1},\cdots,\mathbf{v}_{M}) \in \mathbb{C}^{M \times M-K}$ contains $M-K$ orthogonal eigenvectors belonging to $\text{Ker}\{\mathbf{S}_M\}$ we have:
\begin{equation}
\mathbf{w}_k^H\mathbf{V}\mathbf{V}^H\mathbf{w}_k=0, \qquad k=1,\cdots,K.
\label{ortho}
\end{equation}
A consequence of \eqref{ortho} is that if $t_j$ denotes the sum of the j\textit{th} diagonal of $\mathbf{V}\mathbf{V}^H$ ($j \in \{-M+1,\cdots,M-1\}$ and $t_0=\text{trace}\{\mathbf{V}\mathbf{V}^H\}$) and if $q(y)$ is the polynomial defined by:
\begin{equation}
q(y)\stackrel{\Delta}{=}\sum_{j=-M+1}^{M-1}{t_{-j}y^j}, \qquad y\in \mathbb{C}
\label{defPoly}
\end{equation}
then the zeros of $q(y)$ exhibit inverse symmetry with respect to the unit circle, and $q(y)$ exactly has $K$ zeros on the unit circle, equal to $\{w_k\}_{k=1,\cdots,K}$ (see \cite{Haykin} for a detailed proof). 

In our Gaussian mixture estimation case, the "signal" matrix $\mathbf{S}_M$ \eqref{defSM} is corrupted with the "perturbation" matrix $\mathbf{P}_M$ \eqref{defPM}. When all the component variances tend to zero (ideal case) the perturbation matrix $\mathbf{P}_M$ tends to a null matrix: using \eqref{defAlpha} and \eqref{defPM} we have:
\begin{eqnarray}
\lim\limits_{\sigma_k \to 0}\alpha_{k,l-j}&=&1, \qquad k=1,\cdots,K \\
\lim\limits_{(\sigma_1,\cdots,\sigma_K) \to (0,\cdots,0)} \mathbf{P}_M&=&\mathbf{0}_{M\times M}. 
\end{eqnarray}
Yet, in the general case, $\mathbf{P}_M$ is not null and unfortunately depends on $w_k$. 
\section{Estimation algorithm}
\label{estimationAlgorithm}
The proposed algorithm for estimating the set of component expectations is based on the eigenvector decomposition of the estimation of $\mathbf{R}_M$ \eqref{defRM}, thus neglecting the effect of the perturbation matrix $\mathbf{P}_M$ \eqref{defPM}. Given a set of $N$ observations $\{z_n\}_{n\in \{1,\cdots N\}}$ the algorithm steps are the following:
\begin{enumerate}
\item define a sampling period $T_e$ using \eqref{Te}
\item estimate $M$\footnote{\label{Mvalue}$M=2K$ seems to be a good choice from our simulations but more investigations are needed to optimize the value of $M$.} samples ($M>K$) of the CF using \eqref{estimationCF}
\item build the matrix $\mathbf{R}_M$ using \eqref{defRM}
\item perform a eigenvector decomposition of $\mathbf{R}_M$
\item construct the matrix $\mathbf{V}=(\mathbf{v}_{K+1},\cdots,\mathbf{v}_M)$ with the $M-K$ eigenvectors associated to the $M-K$ smallest eigenvalues of $\mathbf{R}_M$
\item calculate the coefficient of $q(y)$ defined in \eqref{defPoly}
\item calculate the roots of $q(y)$, keep the roots inside the unit circle then identify the $K$ roots that are closest to the unit circle, call them $\{\hat{w}_k\}_{k=1,\cdots,K}$
\item derive $\{\hat{a}_k\}_{k=1,\cdots,K}$ from $\{\hat{w}_k\}_{k=1,\cdots,K}$ using \eqref{ak_from_wk}
\end{enumerate}
\section{Simulation}
In our simulations several types of a six-component Gaussian mixture have first been considered. The set of expectations is equal to $(0,1,2,4,5,6)$, with a difference of one or two between two successive component expectations. Four cases have been studied: common variance and common weight (scenario 1), different variances and common weight (scenario 2), common variance and different weights (scenario 3) and different variances and different weights (scenario 4). A summary of the scenarios is given in Table 1. The parameter $\sigma$ in Table 1 enables to simulate different overlapping situation. The number $N$ of observations per simulation run is 200. The $\mathbf{R}_M$-based algorithm has been run as described in section \ref{estimationAlgorithm} with $M=2K^{\ref{Mvalue}}$. This algorithm has been compared to the EM algorithm \cite{Dempster} with a uniform random start and a maximal number of $100$ iterations. See \cite{McLachlan} for a detailed description of the Gaussian mixture estimation with EM. A constrained version of the EM (EM$_c$) which imposes a common variance and a common mixing weight has been used to prevent the convergence to components with an almost null variance. In all the scenario, EM$_c$ provides better estimates than the standard EM, even in the scenario where the component variances or the component weights are different. Therefore only the performances of EM$_c$ are presented here.
\begin{table}
\begin{center}{
\begin{tabular}{|c|c c|c c|c c|c c|}
\cline{2-9}
\multicolumn{1}{c|}{} & \multicolumn{2}{c|}{ \scriptsize{scenario 1} } & \multicolumn{2}{c|}{\scriptsize{scenario 2}} & \multicolumn{2}{c|}{\scriptsize{scenario 3}} & \multicolumn{2}{c|}{\scriptsize{scenario 4}} \\[1mm]
\hline
\scriptsize{mean} & \scriptsize{var.} & \scriptsize{weight} & \scriptsize{var.} & \scriptsize{weight} & \scriptsize{var.} & \scriptsize{weight} & \scriptsize{var.} & \scriptsize{weight} \\[1mm]
\hline
$0$ & $\sigma^2$ & $\frac{1}{6}$ & $\sigma^2$ & $\frac{1}{6}$ & $\sigma^2$ & $0.2$ & $\sigma^2$ & $0.2$ \\[1mm]
\hline
$1$ & $\sigma^2$ & $\frac{1}{6}$ & $\frac{\sigma^2}{2}$ & $\frac{1}{6}$ & $\sigma^2$ & $0.2$ & $\frac{\sigma^2}{2}$ & $0.2$ \\[1mm]
\hline
$2$ & $\sigma^2$ & $\frac{1}{6}$ & $\sigma^2$ & $\frac{1}{6}$ & $\sigma^2$ & $0.1$ & $\sigma^2$ & $0.1$ \\[1mm]
\hline
$4$ & $\sigma^2$ & $\frac{1}{6}$ & $\frac{\sigma^2}{2}$ & $\frac{1}{6}$ & $\sigma^2$ & $0.2$ & $\frac{\sigma^2}{2}$ & $0.2$ \\[1mm]
\hline
$5$ & $\sigma^2$ & $\frac{1}{6}$ & $\sigma^2$ & $\frac{1}{6}$ & $\sigma^2$ & $0.2$ & $\sigma^2$ & $0.2$ \\[1mm]
\hline
$6$ & $\sigma^2$ & $\frac{1}{6}$ & $\frac{\sigma^2}{2}$ & $\frac{1}{6}$ & $\sigma^2$ & $0.1$ & $\frac{\sigma^2}{2}$ & $0.1$ \\[1mm]
\hline
\end{tabular}}
\end{center}
\caption{Means, variances, weights of the simulated mixture}
\end{table}
To get rid of the permutation ambiguity, the estimation performance is evaluated as follows: If $\textbf{a} \in \mathbb{R}^K$ is the vector of the true component expectations and $\hat{\textbf{a}}_r \in\mathbb{R}^K$ is the vector of the estimated component expectations at simulation run $r$, the performance criterion $e_r$ is defined as the maximal absolute distance between the true and estimated ordered vector of component expectations: 
\begin{equation}
e_r=\left\| \text{sort}(\textbf{a})-\text{sort}(\hat{\textbf{a}}_r)\right\|_{\infty}, \notag
\end{equation}
where $\text{sort}(\textbf{x})$ is the ordered permutation of \textbf{x} and $\left\| \cdot \right\|_{\infty}$ is the infinity norm in $\mathbb{R}^K$. 

The simulation results are presented in Figure 1 for different values of $\sigma$. When $\sigma$ is greater than $0$, there is a risk that the constrained EM converges to a wrong set of estimated component expectations. Typically one estimated component expectation is located in the middle of two true component expectations. For instance, for $\sigma=0.1$, EM$_c$ provides a good set of estimates for only $40\%$ of the run. On the contrary, the $\mathbf{R}_M$-based algorithm provides a perfect set of estimates if $\sigma\leq0.1$ and $e_r$ remains less than $0.2$ if $\sigma<0.2$. In all the investigated scenario and for all the values of $\sigma$, the proposed method outperforms the EM$_c$ algorithm.
\begin{figure}{
\begin{center}
{
\begin{subfigure}{
\includegraphics[height=0.62\linewidth]{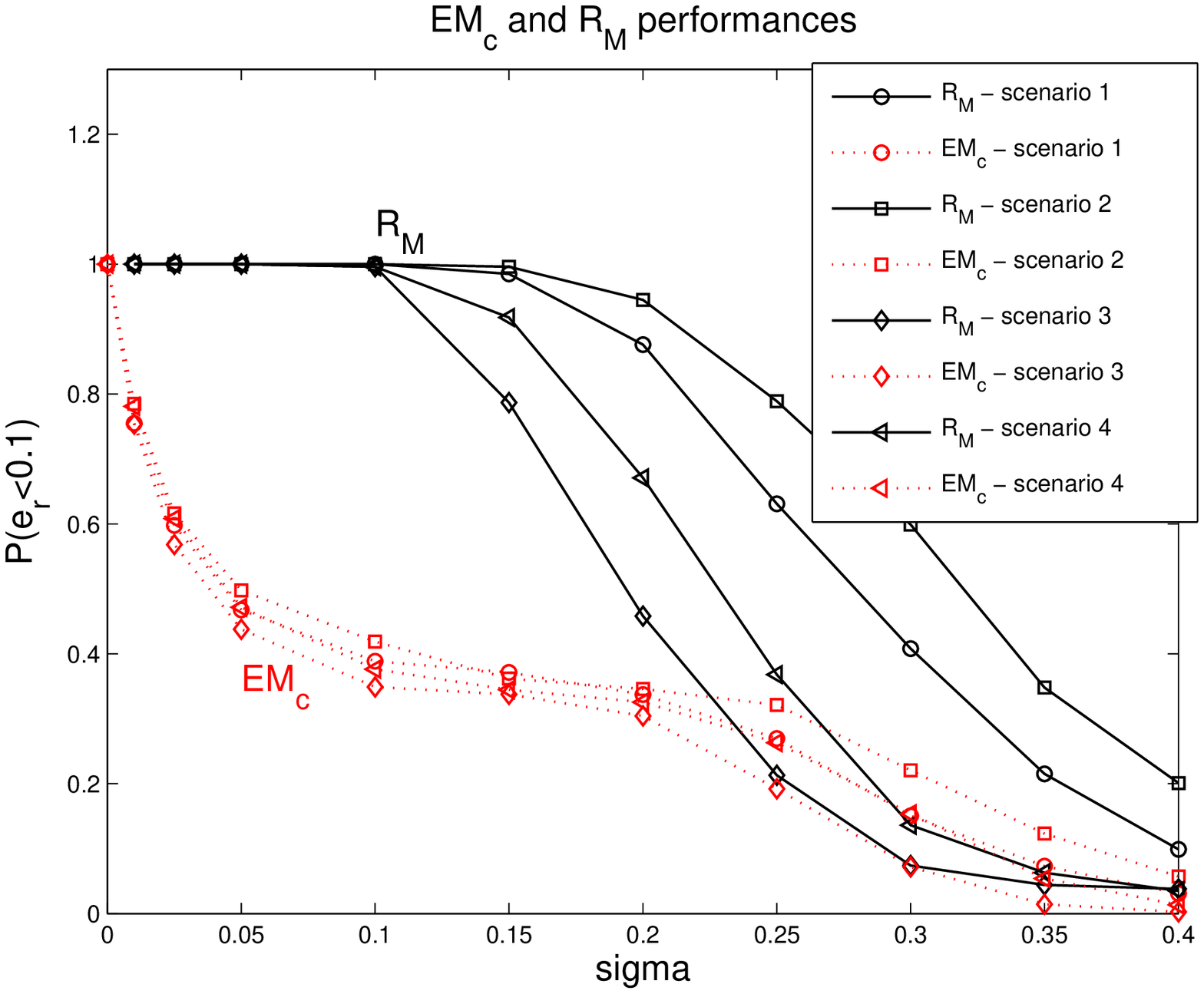}}
\end {subfigure}
\begin{subfigure}{
\includegraphics[height=0.6\linewidth]{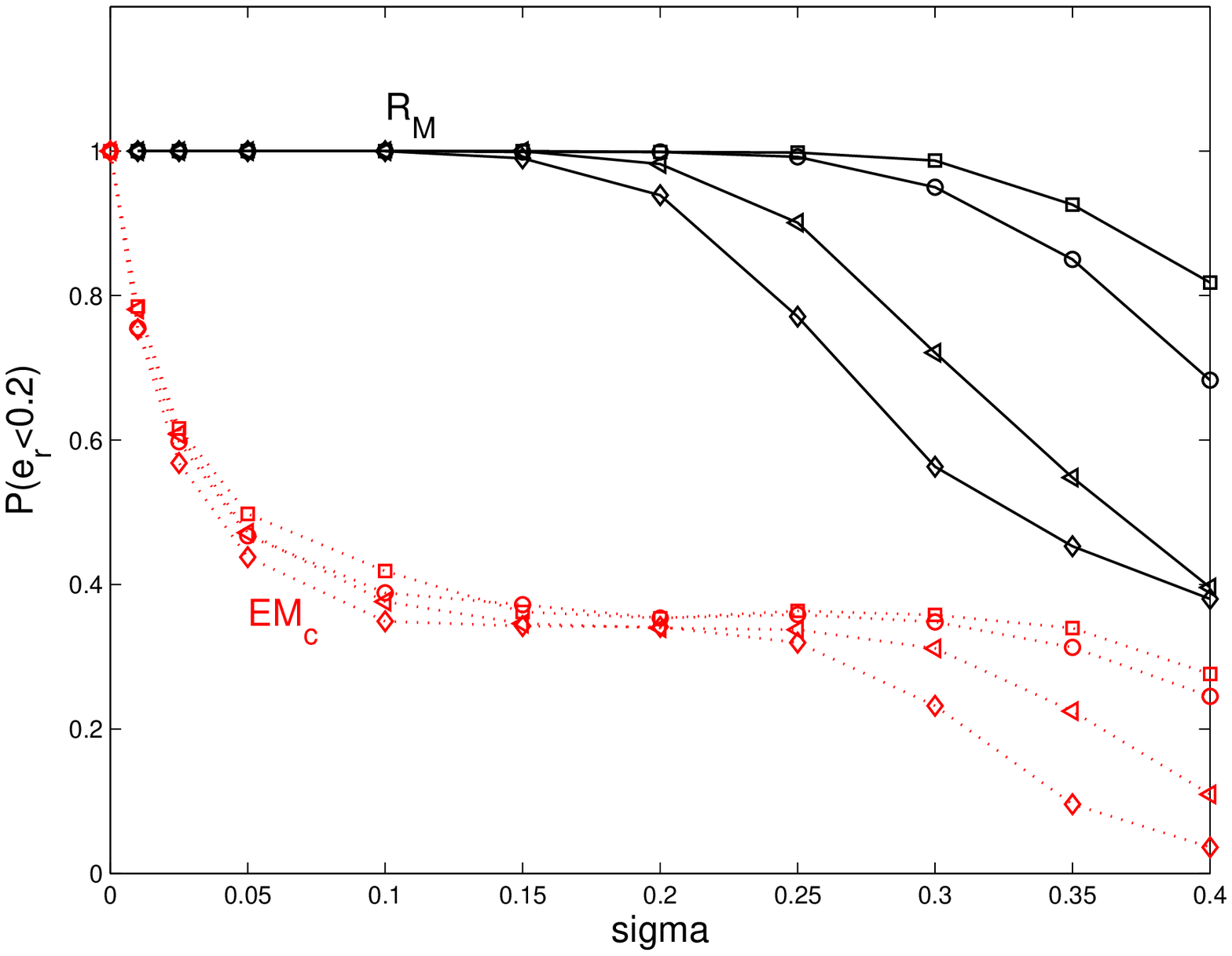}}
\end {subfigure}
}
\end{center}
}
\caption{Performances of the constrained EM (EM$_\text{c}$, dotted lines) and $\mathbf{R}_M$ based algorithm with $M=2K=12$ (full lines) for different values of $\sigma$. For each value of $\sigma$ and for each scenario 10000 simulation runs have been performed. The performance criteria are the probabilities for $e_r$ to be smaller than 0.1 (top), and to be smaller than 0.2 (bottom).}
\end{figure}

\section{Conclusion}
Given a set of observations originating from a $K$-component univariate mixture, we focused on the estimation of the component expectations when the number $K$ of components is known. We proposed a method based on the eigenvector decomposition of a Toeplitz matrix $\mathbf{R}_M$ built from some estimated samples of the first characteristic function. Simulations illustrated the superiority of the proposed method compared with the Expectation-Maximization algorithm on various configurations of a six-component Gaussian mixture. More theoretical investigations are now needed to study the influence of the perturbation matrix \eqref{defPM} on the performances.  

Our current research also deals with the case of an unknown number of components. In figure 2 we plot the eigenvalues of $\mathbf{R}_M$ with $M=10$ obtained in scenario 4 of Table 1 (where the mixture components have different weights and variances) with $N=200$ observations and $\sigma=0.15$. One can see that $K=6$ eigenvalues are clearly greater than 0 while the $M-K=4$ other eigenvalues are almost null. In general, one can therefore expect the eigenvalue decomposition of the matrix $\mathbf{R}_M$ to provide relevant information on the number of components in an observed mixture.

\begin{figure}{
\begin{center}
{\includegraphics[height=0.5\linewidth,width=0.6\linewidth]{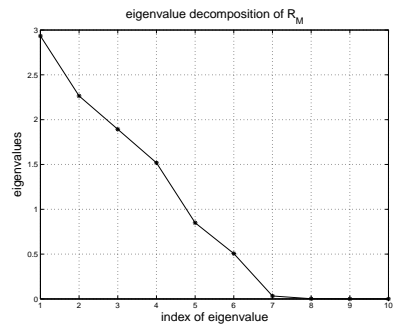}
}\end{center}}
\caption{eigenvalue decomposition of the matrix $\mathbf{R}_M$ with $M=10$ in scenario 4 ($6$ mixture components) for $\sigma=0.15$.}
\end{figure}



\bibliographystyle{IEEEbib}
\bibliography{biblioKfreq}


\end{document}